\documentclass{jetpl}
\twocolumn

\usepackage{graphics,color,epsfig}

\lat

\title{Magnetic ground state and multiferroicity in BiMnO$_3$}

\rtitle{Magnetic ground state and multiferroicity in BiMnO$_3$}

\sodtitle{Magnetic ground state and multiferroicity in BiMnO$_3$}

\author{I.\,V.\,Solovyev$^{+*}$\/\thanks[1]{e-mail: SOLOVYEV.Igor@nims.go.jp},
Z.\,V.\,Pchelkina$^*$\/\thanks[2]{e-mail: pzv@ifmlrs.uran.ru}}

\rauthor{I.\,V.\,Solovyev, Z.\,V.\,Pchelkina}

\sodauthor{Solovyev, Pchelkina}

\address{$^+$Computational Materials Science Center, National Institute for Materials Science, 1-2-1 Sengen, Tsukuba, Ibaraki 305-0047, Japan\\~\\
$^*$Institute of Metal Physics, Russian Academy of
Sciences-Ural Division, 620041 Yekaterinburg GSP-170, Russia}

\dates{Today}{*}

\abstract{
We argue that the centrosymmetric $C2/c$ symmetry in BiMnO$_3$ is
spontaneously broken by antiferromagnetic (AFM) interactions existing
in the system. The true symmetry is expected to be $Cc$,
which is
compatible with the noncollinear
magnetic ground state, where the ferromagnetic order along one crystallographic
axis coexists with the the hidden AFM order
\emph{and related to it ferroelectric polarization}
along two other axes.
The $C2/c$ symmetry can be restored by the
magnetic field $B \sim 35$ Tesla, which switches off the
ferroelectric polarization. Our analysis is based on the
solution of the
low-energy model constructed for the $3d$-bands of BiMnO$_3$,
where all the parameters have been derived from the first-principles
calculations. Test calculations for
isostructural BiCrO$_3$ reveal an excellent agreement with
experimental data.
}


\begin{document}

\maketitle

   BiMnO$_3$ is regarded as one of the prominent multiferroic materials, where the
ferromagnetic (FM) magnetization is coupled to the ferroelectric polarization, thus, giving a
possibility to control the magnetic properties by applying the electric field and
vise versa. The ferromagnetism of BiMnO$_3$ is well established
today: the Curie temperature is about 100 K
and the largest reported magnetization
is $3.92~\mu_B$ per formula unit \cite{belik_mn_07},
which is close to $4~\mu_B$ expected for the fully saturated FM state.
The ferroelectric hysteresis loop was also observed in polycrystalline and thin film samples
of BiMnO$_3$ \cite{SantosSSC},
although the measured polarization was small (about $0.043$ $\mu$C/cm$^2$).
The multiferroic behavior of BiMnO$_3$ is typically attributed to
the existence of two different sublattices: the stereochemical activity of
Bi($6s^2$) lone pairs is believed to be the origin of the structural distortions
(having the $C2$ symmetry) and the inversion symmetry
breaking, which gives rise to the ferroelectricity, while the Mn-sublattice is
responsible for the magnetism~\cite{SeshadriHill}. This point of view dominated over several years
and was documented in many review articles~\cite{Cheong}.

  Nevertheless, in 2007 Belik {\it et al.} have reexamined the crystal structure
of BiMnO$_3$ and argued that below 770 K,
it is best described by the space group $C2/c$,
which has the inversion symmetry \cite{belik_mn_07}.
This finding was later confirmed by Montanari {\it et al.}~\cite{montanari_07}.
According to this new crystal structure information,
BiMnO$_3$ can be only in an \emph{antiferroelectric} state~\cite{spaldin_07} --
the situation, which is rather common
for many distorted perovskites and hardly interesting from the
practical point of view. Of course, these experimental
works raised many
questions. Particularly, what is the origin of the inversion symmetry breaking (if any)
and ferroelectric response in BiMnO$_3$?

\begin{figure}[h!]
\begin{center}
\resizebox{!}{6.2cm}{\includegraphics{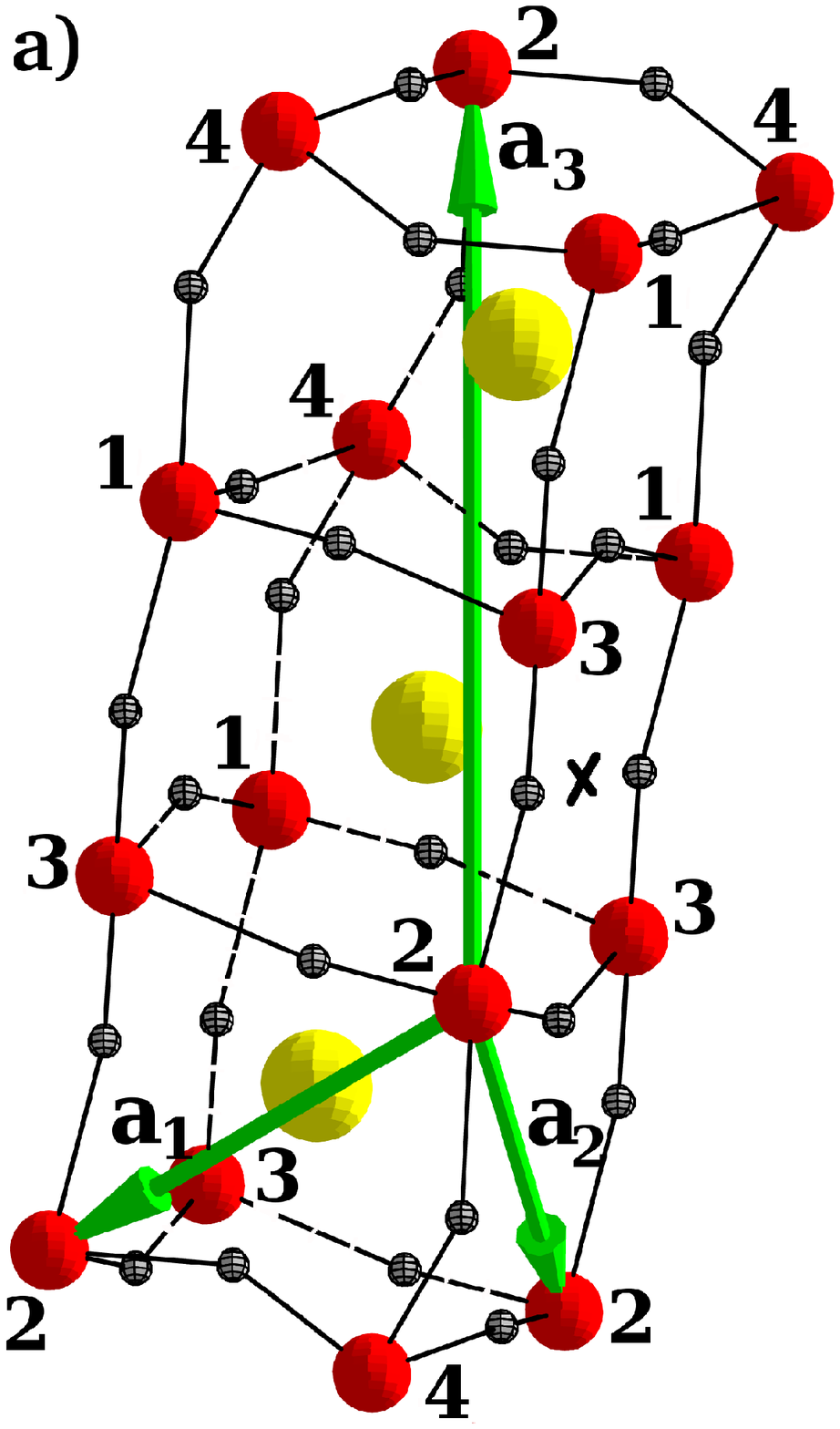}}
\resizebox{!}{6.2cm}{\includegraphics{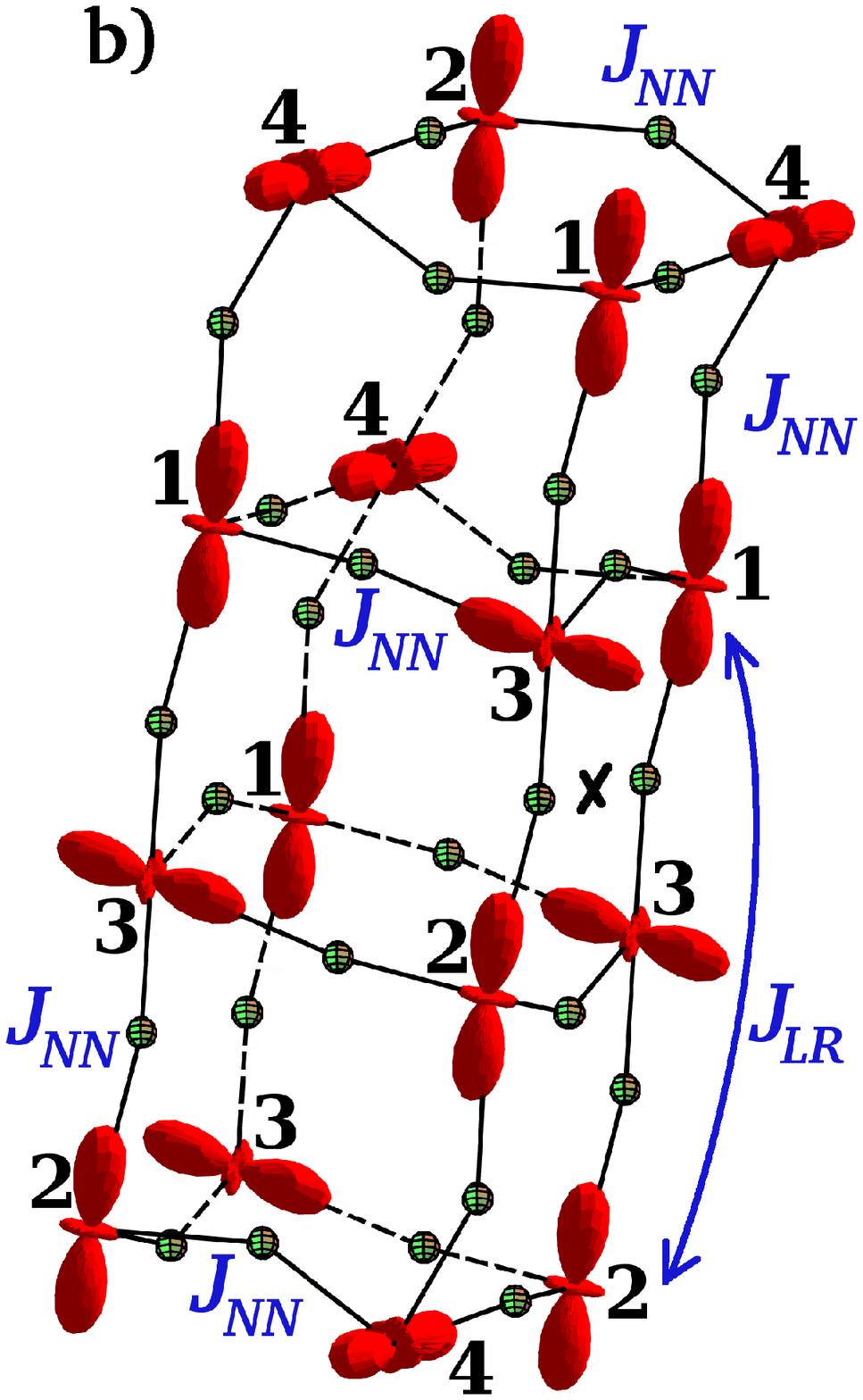}}
\end{center}
\caption{\label{fig.structure}
Fig.1. (a) Crystal structure of BiMnO$_3$.
Bi-, Mn-, and O-atoms are indicated by big, medium, and small spheres,
respectively. Primitive translations are shown by arrows.
Four Mn-atoms, composing the primitive cell of BiMnO$_3$,
are indicated by numbers.
(b) Distribution of charge densities
associated with the occupied $e_g$ orbitals (the orbital ordering)
realized in the low-temperature phase of BiMnO$_3$
and related to it nearest-neighbor ferromagnetic ($J_{NN}$) and
longer range antiferromagnetic ($J_{LR}$)
interactions. The position of the inversion center in the $C2/c$ structure
is marked by the cross.}
\end{figure}

  In our previous work,
we have constructed an effective
Hubbard-type model
for the $3d$-bands of BiMnO$_3$
located near the Fermi level~\cite{solovyev_08}.
All parameters of this model were derived in an \emph{ab initio} fashion
on the basis of
first-principles electronic structure calculations and, apart from
approximations inherent to the
construction of the low-energy model \cite{PRB06a,solovyev_rev},
no adjustable parameters have been used.
Then, from the solution of this model
in the mean-field Hartree-Fock (HF) approximation
we derived parameters of interatomic magnetic interactions
and argued that,
in the $C2/c$ structure, the nearest-neighbor FM interactions
($J_{NN}$) compete with
the longer range antiferromagnetic (AFM)
interactions ($J_{LR}$, see Fig.\ref{fig.structure}).
The latter make the sites 1 and 2 inequivalent
and tend to break the inversion center located in the
midpoint of the cube diagonal connecting these
two sites. Both interactions are
directly related to the orbital ordering realized in BiMnO$_3$ below 550 K.
This suggests that the ferroelectric behavior of BiMnO$_3$
could be related to
some hidden AFM order, which is driven by $J_{LR}$ and which
in~\cite{solovyev_08} was denoted
as the $\uparrow$$\downarrow$$\downarrow$$\uparrow$ order, referring to
the directions of spins at four Mn-atoms in the primitive cell.
But, now we face a more fundamental problem:
how to reconcile the AFM structure,
which is needed in order to break the inversion symmetry,
with the FM behavior of BiMnO$_3$,
which is clearly seen in the experiment?

  The clue to this problem may be related to the
quasi-degeneracy of the FM and
$\uparrow$$\downarrow$$\downarrow$$\uparrow$ AFM states.
Indeed, according to the HF calculations, the total energy
difference between these two states was only
0.5 meV per one formula unit \cite{solovyev_08}.
Although the value itself
maybe at the edge of accuracy of the model calculations,
it gives a clear ideas about characteristic energy scale we have to deal with
in the process of analysis of the magnetic properties of BiMnO$_3$.

  In the present work we further develop this
idea and argue that the AFM $\uparrow$$\downarrow$$\downarrow$$\uparrow$
and FM states can be easily mixed by the
relativistic spin-orbit (SO) interaction $\xi \hat{\bf s} \cdot \hat{\bf l}$.
Although the expectation value of the orbital magnetic moment $\mu_B\hat{\bf l}$
is small in manganites (typically, few hundredths of $\mu_B$~\cite{PRL96}),
the constant of the SO interaction $\xi$ is about 50 meV. Thus, the energy gain caused by the
SO interaction can become comparable with the total energy difference between the FM
and $\uparrow$$\downarrow$$\downarrow$$\uparrow$ AFM states.
First, using general symmetry considerations,
we argue that the true magnetic
ground state of BiMnO$_3$ is noncollinear and includes both FM and AFM components
of the magnetic moments. This naturally resolves
the puzzle of multiferroicity of BiMnO$_3$: the inversion symmetry is broken by the
hidden AFM order, while the FM component appears due to the canting
of the magnetic moments.
The idea will be further supported by the direct optimization of the magnetic structure
of BiMnO$_3$ in the model HF calculations with the SO interaction.
Finally, we show how the ferroelectric polarization can be controlled
by the external magnetic field, which acts on the FM component
of the magnetic moments. Particularly, we predict the
existence of the critical field $B_c$, which restores the inversion
symmetry and switches off the ferroelectric polarization.

  We begin with the symmetry considerations.
The space group $C2/c$ has four symmetry operations:
$\hat{S}_1$$=$$\hat{E}$,
$\hat{S}_2$$=$$\hat{I}$,
$\hat{S}_3$$=$$\{ \hat{m}_y|{\bf a}_3/2 \}$,
$\hat{S}_4$$=$$\{ \hat{C}^2_y|{\bf a}_3/2 \}$,
where
$\hat{E}$ is the unity, $\hat{I}$ is the inversion, $\hat{m}_y$ is the mirror
reflection of the $y$-axis, and $\hat{C}^2_y$ is the $180^\circ$ rotation around the $y$-axis.
The last two operations are combined with the translation by the vector ${\bf a}_3/2$
(see Fig.\ref{fig.structure}a).
The directions of $x$, $y$, and $z$ are related to the
directions of the primitive translations
as follows: $x$$\|$$({\bf a}_1$$+$${\bf a}_2$$+$$0.34{\bf a}_3)$, $y$$\|$$({\bf a}_2$$-$${\bf a}_1)$, and
$z$$\|$${\bf a}_3$.
Four Mn-atoms are divided in two classes:
(1,2) and (3,4). Atoms of one class are
transformed by the symmetry operations
only to each other, but not to atoms of another class.
The magnetic groups corresponding to the $C2/c$ space group
are obtained by combining the symmetry operations $\hat{S}_2$-$\hat{S}_4$
with the time inversion $\hat{T}$,
which additionally flips the directions of the magnetic moments.
Then, there are two magnetic groups, which can be
regarded as candidates
for the magnetic ground state of BiMnO$_3$:
${\bf G}_1$$=$$\{ \hat{S}_1, \hat{S}_2, \hat{S}_3, \hat{S}_4 \}$ and
${\bf G}_2$$=$$\{ \hat{S}_1, \hat{S}_2, \hat{S}_3\hat{T}, \hat{S}_4\hat{T} \}$.
The third possibility includes the symmetry operation
$\hat{S}_2\hat{T}$$\equiv$$\hat{I}\hat{T}$. However, since $\hat{I}$ transforms the
atoms 3 and 4 to themselves, the symmetry operation $\hat{I}\hat{T}$
corresponds to the nonmagnetic sublattice (3,4).
Such a situation, although formally possible,
cannot be realized as the magnetic ground state of BiMnO$_3$, because it
contradicts to the first Hund rule \cite{solovyev_08}.
Then, $x$ and $z$ projections of the magnetic moments
obey the same transformation rules under the symmetry operations
$\hat{S}_1$-$\hat{S}_4$.
Therefore,
the magnetic structure of BiMnO$_3$ can be generally
abbreviated as ${\cal AB}$-${\cal CD}$, where each capital letter stands
for the magnetic arrangement (ferromagnetic - $F$, antiferromagnetic - $A$,
or with zero projection of the magnetic moment - $Z$) formed by the
$x$ and $y$ projections of the magnetic moments of the atoms (1,2)
and (3,4): first two capital letters describe the magnetic
arrangement in the pair (1,2) formed by the $x$ (${\cal A}$)
and $y$ (${\cal B}$) projections of the magnetic moments; while second
two capital letters describe the magnetic
arrangement in the pair (3,4) formed by the $x$ (${\cal C}$)
and $y$ (${\cal D}$) projections of the magnetic moments.
For example, the magnetic structure corresponding to ${\bf G}_1$ and ${\bf G}_2$
can be denoted as $ZF$-$AF$ and $FZ$-$FA$, respectively. Moreover, we
consider the solutions with the spontaneous symmetry breaking
corresponding to three possible subgroups of the space group $C2/c$.
From this point of view, the most promising are two magnetic subgroups:
${\bf G}_3$$=$$\{ \hat{S}_1, \hat{S}_3 \}$
(the space group $Cc$, No. 9 in International Tables)
and
${\bf G}_4$$=$$\{ \hat{S}_1, \hat{S}_3\hat{T} \}$,
which preserve the atomic classes (1,2) and (3,4).
They correspond
to the magnetic structures $AF$-$AF$ and $FA$-$FA$, respectively.
In the other words, contrary to ${\bf G}_1$ and ${\bf G}_2$,
the magnetic subgroups ${\bf G}_3$ and ${\bf G}_4$ allow for
the nontrivial AFM arrangement in the pair (1,2) with finite
projections of the magnetic moments. Such a situation becomes possible due to the breaking
of the inversion symmetry $\hat{I}$
(relative to the center, which is shown by the cross in Fig.\ref{fig.structure}b).
In the case of the collinear magnetic arrangement,
the magnetic structures $AF$-$AF$ and $FA$-$FA$
are reduced to either
FM or
AFM $\uparrow$$\downarrow$$\downarrow$$\uparrow$ ones,
which have been considered in \cite{solovyev_08}
without the SO interaction.
Other magnetic subgroups, generated by
$\hat{S}_2$$=$$\hat{I}$ and $\hat{S}_4$$=$$\{ \hat{C}^2_y|{\bf a}_3/2 \}$
destroy the atomic classes and make the atoms forming the pairs (1,2) and
(3,4) inequivalent.
Although such solutions are formally possible, all of them are
unstable and not realized as the magnetic ground state of BiMnO$_3$,
as it will become clear from the solution of the low-energy model.

\begin{figure}[h!]
\begin{center}
\resizebox{!}{3.5cm}{\includegraphics{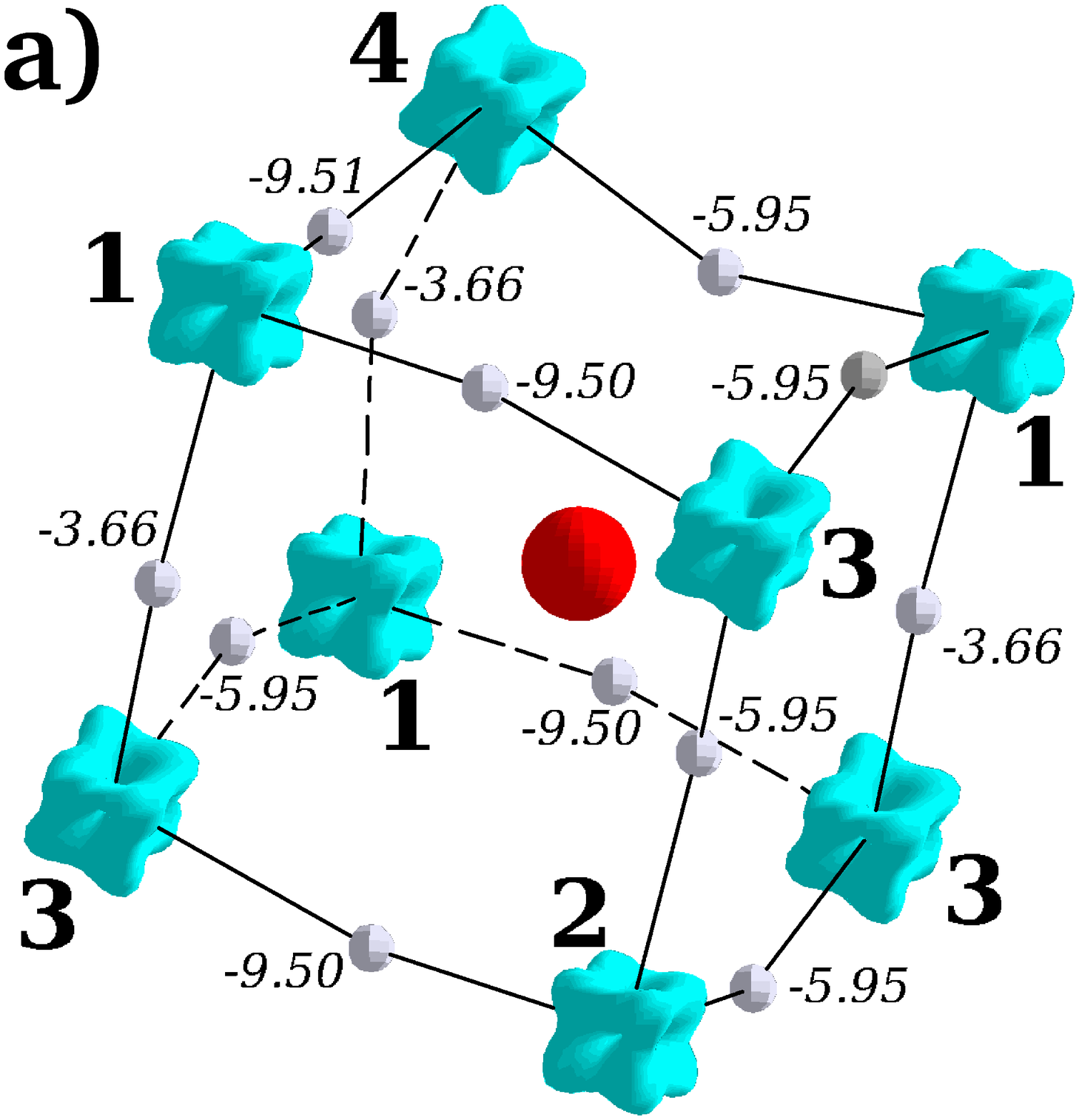}}
\resizebox{!}{3.5cm}{\includegraphics{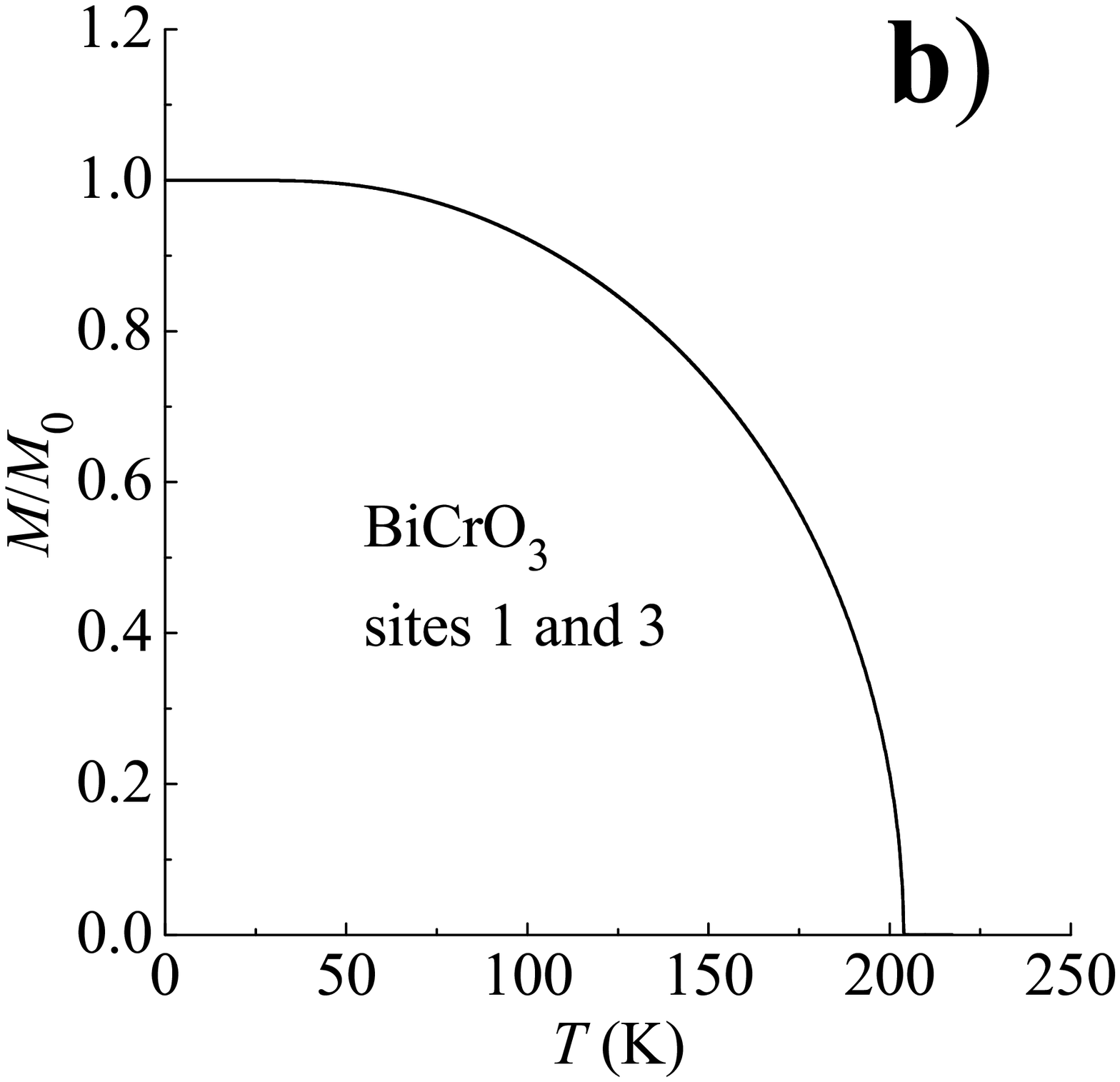}}
\end{center}
\caption{\label{fig.bicro3}
Fig.2. (a) Distribution of occupied $t_{2g}$-orbitals
at the Cr-sites
and corresponding parameters of Heisenberg model ($J_{ij}S^2$, where $S$$=$$3/2$)
associated with different bonds in BiCrO$_3$. Bi- and O-atoms are shown by big and small spheres,
respectively. (b) Corresponding temperature dependence of the local magnetization
at the sites 1 and 3 in the mean-field approximation, which demonstrates
that the two curves are practically indistinguishable,
despite the fact that the sites 1 and 3 are inequivalent and belong to different
atomic classes.}
\end{figure}

  Our next goal is the search for the true (and so far unknown)
magnetic ground state of BiMnO$_3$
based on the solution of the low-energy model.
However, before doing any predictions for BiMnO$_3$, we would like to show some test
calculations for the isostructural compound BiCrO$_3$~\cite{belik_cr}, which give
some idea about the accuracy of our approach.
The formal configuration of Cr$^{3+}$ ions in BiCrO$_3$ is $t_{2g}^3$. Since the
$t_{2g}$-levels are well separated form the $e_g$-ones by the crystal-field (CF) effects, the orbital
degrees of freedom in BiCrO$_3$ are quenched and the Hubbard model can be further
mapped onto the Heisenberg model
$\hat{\cal H}_S = -$$\sum_{i>j}J_{ij}\hat{\bf S}_i \hat{\bf S}_j$
with the spin $3/2$.\footnote{We used the same strategy as in our
previous work on BiMnO$_3$~\cite{solovyev_08}. First, we construct an
effective
Hubbard-type model. The main difference of the model parameters from the
ones reported for BiMnO$_3$~\cite{solovyev_08} is the following:
the CF-splitting
between two $e_g$-levels is small
(134 and 185 meV for the sites 1 and 3, respectively),
while the $t_{2g}$-$e_g$ splitting is large
(about 2.1 eV). Moreover, the on-site Coulomb repulsion $U$ is larger in
BiCrO$_3$: 2.63 and 2.73 eV for the sites 1 and 3, resectively.
Then, we calculate $J_{ij}$ using infinitesimal rotations
near the $\uparrow$$\uparrow$$\downarrow$$\downarrow$ (or G-type) AFM ground state.
In BiCrO$_3$, other magnetic configurations yield practically
the same set of the exchange parameters $J_{ij}$, meaning that the mapping onto the
Heisenberg model is universe and does not depend on the magnetic state.
}
The parameters of such a model
are shown in Fig.\ref{fig.bicro3}a.
Generally, the distribution of the magnetic moments
associated with the existence of two
inequivalent sublattices (1,2) and (3,4)
should be nonuniform. The magnitude of this effect can be estimated
in the mean-field approximation. However, it appears that
from the magnetic point of view,
the sublattices (1,2) and (3,4) are nearly equivalent. Indeed,
the behavior of local magnetization at the sites 1 and 3 (Fig.\ref{fig.bicro3}b)
is practically
indistinguishable in the whole temperature range below the N\'{e}el temperature
$T_N$ (about 204 K in the mean-field approximation). Then, a proper
(and much better) estimate for $T_N$
can be obtained by assuming that the main excitations in BiCrO$_3$
can be described by (uniform) spin waves and employing the
renormalized spin-wave theory, which takes into account the
spacial correlations between the spins \cite{Tyablikov}. It yields $T_N$$=$$123$ K, which is
very close to the experimental
value $T_N$$=$$109$ K \cite{belik_cr_mag}.

\begin{figure}[h!]
\begin{center}
\resizebox{7cm}{!}{\includegraphics{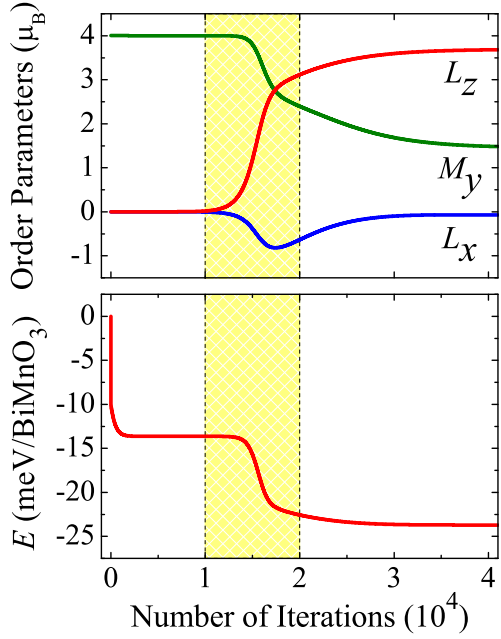}}
\end{center}
\caption{\label{fig.iter} Fig.3. Convergence of the ferromagnetic moment
$M_y \equiv M_y^1 = M_y^2$, two antiferromagnetic order parameters
$L_x = (M^1_x$$-$$M^2_x)/2$ and $L_z = (M^1_z$$-$$M^2_z)/2$,
and the total energy in the Hartree-Fock
calculations after switching on the spin-orbit interaction. The
initial configuration corresponded to the self-consistent
Hartree-Fock solution for the ferromagnetic state without
spin-orbit interaction, where
all magnetic moments were aligned along the $y$-axis.
The shaded area is the region corresponding to the change of the
symmetry ${\bf G}_1 \Rightarrow {\bf G}_3$
of the Hartree-Fock solution
by the antiferromagnetic interactions.}
\end{figure}

  Being encouraged by these results, we now turn to
the problem of the magnetic ground state and
the origin of the ferroelectric behavior of BiMnO$_3$.
First, we perform the delicate optimization of the magnetic
structure of BiMnO$_3$ based on the HF solution of the low-energy
model with the SO interaction. The parameters of the low-energy
model were reported in~\cite{solovyev_08}. The solution of the
low-energy model with the SO interaction
was performed along the same line as for other distorted
perovskite oxides considered in the
review article \cite{solovyev_rev}.
A typical iterative procedure is shown in Fig.\ref{fig.iter}.
In this case, we start with the FM solution, where all
magnetic moments were aligned along the $y$-axis,
switch on the SO interaction, and monitor the behavior of the
FM moments ($M^1_y$$=$$M^2_y$) as well as two AFM order parameters
$L_x$$=$$(M^1_x$$-$$M^2_x)/2$ and $L_z$$=$$(M^1_z$$-$$M^2_z)/2$,
which were developed
at
the sites 1 and 2 in the process of
iterative solution of the HF equations. In these notations,
$M^i_a$ is the projection of the magnetic moment at the site $i$
parallel to the $a$-axis, where $a$$=$ $x$, $y$, or $z$.
In the process of first approximately $10^4$ iterations, the solution obeys the
${\bf G}_1$ symmetry, and the iterative procedure is
accompanied by the small but steady decrease of the
total energy (following the sharp drop right after switching
on the SO interaction). As expected, the $x$- and $z$-projections
of the magnetic moments at the sites 1 and 2
are exactly equal to zero in this region
(so as the AFM order parameters $L_x$ and $L_z$).
However, in the process of next $10^4$ iterations
we clearly observe the
lowering of the magnetic symmetry
${\bf G}_1 \Rightarrow {\bf G}_3$, which is accompanied
by the growth of AFM order parameters $L_x$ and $L_z$
(at the expense of
the FM magnetization along $y$) and the step decrease of the
total energy of the system (by about 10 meV per one formula unit).
This solution finally converges to the following values of
the magnetic moments ${\mathbf M}$$=$$(M_x,M_y,M_z)$ at the sites 1-4 (in $\mu_B$):
\begin{eqnarray*}
{\mathbf M}^1 = (\phantom{-}0.08,-1.45,          -3.69) \nonumber \\
{\mathbf M}^2 = (          -0.08,-1.45,\phantom{-}3.69) \nonumber \\
{\mathbf M}^3 = (          -0.97,-2.02,          -3.27) \nonumber \\
{\mathbf M}^4 = (\phantom{-}0.97,-2.02,\phantom{-}3.27) \nonumber
\end{eqnarray*}
We also tried to start with other magnetic configurations
(including all magnetic configurations considered in \cite{solovyev_08} with
different directions of the magnetic moments). However, all of them finally
converged to the solution described above. For example, by starting with
the AFM $\uparrow$$\downarrow$$\downarrow$$\uparrow$ configuration and
aligning all magnetic
moments along $y$-axis, one can obtain a metastable solution having
the ${\bf G}_2$ symmetry,
which is only 3 meV higher that the ${\bf G}_3$ solution discussed above.
However, even this metastable solution finally converged to the
${\bf G}_3$ one, which seems to be the true global minimum of the
total energy of the system. These calculations clearly show the
advantages of working with the low-energy models, because in order to reach
the true global minimum with the SO interaction, we typically need
a huge number of iterations, which is only affordable in the model approach.

  We would also like to emphasize that
the symmetry of the magnetic ground state, obtained
in the present work, is incompatible with the noncentrosymmetric
crystal structure $C2$, which was proposed in earlier studies~\cite{atou_99} and
was shown to be unstable~\cite{spaldin_07} with respect to the
centrosymmetric
$C2/c$ structure~\cite{belik_mn_07}.
Instead, we propose that, at least below
the magnetic transition temperature, the crystal structure of BiMnO$_3$ should have the $Cc$ symmetry.
This finding should be checked experimentally and we hope that our work
will stimulate further activities in this direction.

\begin{figure}[h!]
\begin{center}
\resizebox{7cm}{!}{\includegraphics{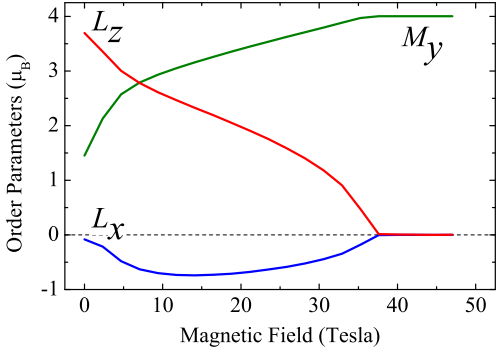}}
\end{center}
\caption{\label{fig.field} Fig.4. Behavior of the ferromagnetic magnetization
$M_y \equiv M_y^1 = M_y^2$ and the antiferromagnetic order parameters
$L_x = (M^1_x$$-$$M^2_x)/2$ and $L_z = (M^1_z$$-$$M^2_z)/2$
in the external magnetic
field along the $y$-axis.}
\end{figure}

  The magnetic group ${\bf G}_3$ allows for the net electric polarization ${\mathbf P}$
in the plane $zx$, which is a normal vector and
does not changes its sign under $\hat{S}_3$$=$$\{ \hat{m}_y|{\bf a}_3/2 \}$.
On the other hand, ${\mathbf M}$ is an axial vector and both AFM order parameters
$L_x$ and $L_z$
change the sign under
$\hat{S}_3$. Therefore, the ferroelectric
polarization caused by the magnetic degrees of freedom
must be at least bilinear with respect to
$L_x$ and $L_z$ \cite{SmolenskiiChupis}:
\begin{equation}
P_a = \sum_{bc} \chi_{abc} L_b L_c,
\label{eqn:polarization}
\end{equation}
where the symbols $a$, $b$, and $c$ denote the $z$ and $x$
projections of the vectors. Since $L_x$ and $L_z$ are related to the
the FM magnetization $M_y$ (due to the conservation of $|{\mathbf M}|$),
the value of ${\mathbf P}$ can be
controlled by the external magnetic
field ${\mathbf B}$$=$$(0,B,0)$, applied along $y$ and coupled to $M_y$.
Fig.\ref{fig.field} shows results of HF calculations in the external magnetic field.
The corresponding interaction term is described by
$\hat{\cal H}_B = -$$\mu_B {\mathbf B} \cdot (2\hat{\bf s}$$+$$\hat{\bf l})$.
As expected, ${\mathbf B}$ saturates the FM magnetization $M_y$. Then,
other two projections of the magnetic moments will decrease,
so as the AFM order parameters $L_x$ and $L_z$.
Therefore, the ferroelectric polarization (\ref{eqn:polarization})
should also decrease.
When the field exceeds some critical value $B_c \sim 35$ Tesla,
both
$L_x$ and $L_z$ vanish. Thus, $B > B_c$ restores the
full symmetry of the system, ${\bf G}_3 \Rightarrow {\bf G}_1$,
and switches off the
ferroelectric polarization.
This finding is very important for practical applications as
it
clearly demonstrates
how the ferroelectric behavior of BiMnO$_3$ can be controlled by the
magnetic field. Of course, the obtained $B_c$ is too large for any practical applications.
Nevertheless, this theoretical value may be overestimated
because the model itself does not include several important ingredients such as
the magnetic polarization of the oxygen sites~\cite{solovyev_08}.
The latter will additionally
stabilize the FM interactions~\cite{solovyev_08,solovyev_JPSJ} and
therefore decrease $B_c$.
On the other hand, the FM interactions at the oxygen sites will be
partly compensated by AFM interactions, which are additionally stabilized
by
correlation effects
beyond the HF approximation~\cite{solovyev_JPSJ}.
Moreover, the relative strength of FM and AFM interactions
(and therefore $B_c$) could be tuned by the external factors such as
pressure,
defects or the lattice mismatch
in the thin films of BiMnO$_3$ deposed on different substrates~\cite{solovyev_08}.

  In conclusion, we predict that the centrosymmetric $C2/c$ symmetry in BiMnO$_3$
is spontaneously broken by hidden AFM interactions, which downgrade
the actual symmetry to the $Cc$ one (No. 9 in International Tables).
The space group $Cc$ has only one nontrivial symmetry
operation (the mirror reflection of the $y$-axis with
subsequent translation by ${\bf a}_3/2$),
which
allows for the existence of both ferromagnetic order
(along the $y$-axis) and ferroelectric polarization (in the plane perpendicular to $y$).
The full $C2/c$ symmetry can be restored in
the external magnetic field, which
can be used
in practical applications
in order
to
completely
switch off the ferroelectric polarization.

The work of IVS is partly supported by Grant-in-Aid for Scientific
Research in Priority Area ``Anomalous Quantum Materials''
and Grant-in-Aid for Scientific
Research (C) No. 20540337
from the
Ministry of Education, Culture, Sport, Science and Technology of
Japan.
The work of ZVP is partly supported
by Dynasty Foundation,
Grants of President of Russia MK-3227.2008.2, and scientific school
grant SS-1929.2008.2.

\end{document}